\documentclass[apjl]{emulateapj}

\usepackage{graphicx}
\usepackage{amsmath}
\usepackage{natbib}

\usepackage{color}

\begin{document} 

\title{
  On the Origin of the sub-Jovian Desert in the Orbital-Period--Planetary-Mass
  Plane
}

\author{Titos Matsakos and Arieh K\"onigl}

\affil{
  Department of Astronomy \& Astrophysics and The Enrico Fermi Institute,
  The University of Chicago, Chicago, IL 60637, USA
}

\shortauthors{Matsakos \& K\"onigl}
\shorttitle{Origin of Sub-Jovian Desert}

\begin{abstract}
Transit and radial velocity observations indicate a dearth of sub-Jupiter--mass
planets on short-period orbits, outlined roughly by two oppositely sloped lines
in the period--mass plane.
We interpret this feature in terms of high-eccentricity migration of planets
that arrive in the vicinity of the Roche limit, where their orbits are tidally
circularized, long after the dispersal of their natal disk.
We demonstrate that the two distinct segments of the boundary are a direct
consequence of the different slopes of the empirical mass--radius relation for
small and large planets, and show that this relation also fixes the mass
coordinate of the intersection point.
The period coordinate of this point, as well as the detailed shape of the lower
boundary, can be reproduced with a plausible choice of a key parameter in the
underlying migration model.
The detailed shape of the upper boundary, on the other hand, is determined by
the post-circularization tidal exchange of angular momentum with the star and
can be reproduced with a stellar tidal quality factor $Q^\prime_*\sim10^6$.
\end{abstract} 


\maketitle

\section{Introduction}
\label{sec:intro}

As has been known for some time now, radial-velocity surveys of exoplanets
exhibit an abrupt drop in the number of hot Jupiters (HJs) for orbital periods
$P_\mathrm{orb}\lesssim3$\,days \citep[e.g.,][]{Cumming+08}, corresponding to a
pileup of such planets near $P_\mathrm{orb}=3$\,days \citep[e.g.,][]{Gaudi+05}
and their paucity at shorter periods \citep[e.g.,][]{ZuckerMazeh02}.
(We define HJs as planets that have masses in the range
$M_\mathrm{p}\sim0.3$--$3\,M_\mathrm{J}$, where $M_\mathrm{J}$ is Jupiter's
mass, and periods $P_\mathrm{orb}\lesssim10$\,days.)
Planet candidates identified by transit measurements exhibit a similar sharp
drop in the number count below $P_\mathrm{orb}\sim3$--$4$\,days for objects with
radii $R_\mathrm{p}\gtrsim4\,R_\earth$ \citep[e.g.,][]{Howard+12,Fressin+13}.
One interpretation of these results invokes planet--planet scattering events
that place one of the interacting planets on a highly eccentric orbit: that
planet thereby attains a small pericenter distance, where its orbit is
circularized through tidal interaction with the host star
\citep[e.g.,][]{FordRasio06}.
In this picture, the shortest possible pericenter distance is given by the Roche
limit $a_\mathrm{R}$, where the planet starts to be tidally disrupted, and the
planet's resulting circular radius must therefore exceed $\sim2\,a_\mathrm{R}$
(where the factor of $2$ follows from conservation of orbital angular momentum).
A planet can also be placed on a highly eccentric orbit by a more gradual
process such as Kozai \citep[e.g.,][]{WuMurray03,FabryckyTremaine07} or secular
\citep[e.g.,][]{WuLithwick11,Petrovich15b} migration.
An alternative interpretation attributes the scarcity of high-$M_\mathrm{p}$
planets on short-period orbits to tidal exchange of angular momentum with the
star, which causes close-in giant planets to spiral inward and get ingested by
their host \citep[e.g.,][]{Jackson+09,TeitlerKoenigl14}.

As data continued to accumulate, more details have emerged about the structure
of the planet distribution at low values of $P_\mathrm{orb}$.
In particular, \cite{SzaboKiss11} considered the
($P_\mathrm{orb}$,~$M_\mathrm{p}$) plane and identified a region (defined by
$P_\mathrm{orb}<2.5$\,days and $M_\mathrm{p}$ between $0.02$ and
$0.8\,M_\mathrm{J}$) with a pronounced dearth of planets, which they termed the
``sub-Jupiter desert.''\footnote{
As discussed in that paper as well as in the other studies of this topic that we
cite, a complementary description of the sub-Jovian desert can be given in the
($P_\mathrm{orb}$,~$R_\mathrm{p}$) plane.
However, in view of the degeneracy of the $M_\mathrm{p}(R_\mathrm{p})$ relation
for large planets (see Equation~(\ref{eq:RM}) below), we consider only the
($P_\mathrm{orb}$,~$M_\mathrm{p}$) plane in this work.}
They noted that this region has both an upper boundary, consisting of HJs, and a
lower boundary, consisting of close-in super-Earths (SEs), but is devoid of hot
Neptunes.
Similar identifications were made by \cite{BeaugeNesvorny13} and by
\cite{Mazeh+16} (with the latter authors approximating the boundary by two
intersecting straight lines in the
($\log{P_\mathrm{orb}}$,~$\log{M_\mathrm{p}}$) plane), and it is generally
accepted that this feature is not the result of some observational
bias.\footnote{
In fact, several planets are known to lie inside the nominal desert, although so
far all such planets are located near the desert's edges
\citep[e.g.,][]{Colon+15}.}
However, the origin of the desert is still being debated
\citep[e.g.,][]{SzaboKiss11,Benitez-Llambay+11,BeaugeNesvorny13,
KurokawaNakamoto14,Batygin+16,Mazeh+16}.
In one proposed scenario \citep{KurokawaNakamoto14}, the desert is attributed to
the evaporation of close-in giant planets by the stellar radiation field,
followed by Roche-lobe overflow.
However, this picture also implies the complete loss of the gaseous envelopes of
the less massive SEs, which is inconsistent with the fact that the inferred
radii of the planets near the lower boundary of the desert are generally well in
excess of $1.6\,R_\earth$ and are thus unlikely to be purely rocky
\cite[see][]{Rogers15}.
In an alternative scenario \citep{Valsecchi+14,Valsecchi+15}, HJs arrive at the
Roche limit and undergo rapid mass loss through Roche-lobe overflow, with
photoevaporation further contributing to the removal of their envelopes.
It was proposed that this mechanism could convert HJs into SEs, but it is not 
obvious that the resulting low-mass planet distribution would be consistent with
the observed shape of the desert's lower boundary.
Yet another possibility is that the desert arises from the in-situ formation of
HJs by gas accretion onto SE cores \citep[][]{Boley+16,Batygin+16}.
However, this picture also faces various challenges
\citep[e.g.,][]{InamdarSchlichting15}.

In this Letter we propose that the boundary of the desert reflects the locus of
the innermost circularized orbits of planets that arrived in the vicinity of the
host star through planet-planet scattering, Kozai migration (arising from an
interaction with either a stellar or a planetary companion), or a secular
process long after the protoplanetary disk had dispersed.
As was already pointed out by \cite{FordRasio06}, the Roche limit $a_\mathrm{R}$
(which is taken to be the distance of closest approach for a planet arriving via
either scattering or Kozai migration) is a function of both $R_\mathrm{p}$ and
$M_\mathrm{p}$ (for a given host mass $M_*$), and it can therefore be expressed
(using the appropriate mass--radius relation) as a function of $M_\mathrm{p}$.
A similar expression can be written down for the expected distance of closest
approach in the secular chaos scenario. Using an empirical
$R_\mathrm{p}(M_\mathrm{p})$ relation, we demonstrate that this picture can
account for the basic shape of the desert's boundary, with its oppositely sloped
upper and lower segments.
We further show that the detailed structure of the boundary can be qualitatively
reproduced when one also takes into account the orbital evolution induced in
close-in planets by their tidal interaction with the host star.
These results on the behavior of late-arriving close-in planets complement
recent inferences about the evolution of an earlier generation of HJs that had
migrated though the natal disk \citep{MatsakosKonigl15}.

\section{Modeling Approach}
\label{sec:model}

Since most of the planets in the vicinity of the desert's boundary are inferred
to have low eccentricities ($e<0.1$; see Figure~\ref{fig:M-P} below), we focus
on scenarios in which planets are placed on high-eccentricity orbits that are
circularized by internal tidal dissipation after the planets arrive in the
vicinity of the host star. For any given system, we label by $t_\mathrm{arr}$
the time at which the planet under consideration arrives at the stellar vicinity
with orbital eccentricity $e_0$.
After arrival, its orbit is circularized to a radius $a_\mathrm{c}$ (where its
orbital period is $P_\mathrm{orb}$) on a timescale
$\tau_\mathrm{c}\propto{Q^\prime_\mathrm{p}P_\mathrm{orb}^{13/3}M_\mathrm{p}
R_\mathrm{p}^{-5}}$, where $Q^\prime_\mathrm{p}$ is the planet's tidal quality
factor \citep[e.g.,][]{Matsumura+10b}.
Planets along the lower boundary of the desert have low values of
$R_\mathrm{p}$, which has the effect of increasing $\tau_\mathrm{c}$.
However, the smallest such planets would be mostly rocky and would thus also
have low values of $Q^\prime_\mathrm{p}$ \citep[e.g.,][]{GoldreichSoter66}, so
their orbits might still be circularized over the planets' typical ages
($t_\mathrm{age}\gtrsim1$\,Gyr). 
In this work we relate $R_\mathrm{p}$ and $M_\mathrm{p}$ through the empirical
relation obtained by \cite{Weiss+13}.
They divided their sample into small and large planets (superscripts S and L,
respectively), separated at $M_\mathrm{p}\approx150\,M_\earth$, and wrote down
power-law relationships that we further approximate here by adopting the median
incident flux of the data set ($F=8.6\times10^8$\,erg\,s$^{-1}$\,cm$^{-2}$) and
rounding to the first decimal figure in each exponent:
\begin{equation}
\frac{R_\mathrm{p}^\mathrm{S}}{R_\mathrm{J}}\sim1.6\,\left(\frac{M_\mathrm{p}}{M_\mathrm{J}}\right)^{1/2}\,,\quad\quad\frac{R_\mathrm{p}^\mathrm{L}}{R_\mathrm{J}}\sim1.5\,\left(\frac{M_\mathrm{p}}{M_\mathrm{J}}\right)^0\,.\label{eq:RM}
\end{equation}

The tidal interaction with the star can also lead to orbital decay and eventual
ingestion by the host.
This occurs on a nominal timescale
$\tau_\mathrm{d}\sim[(Q^\prime_*/Q^\prime_\mathrm{p})(M_*/M_\mathrm{p})^2
(R_\mathrm{p}/R_*)^5]\,\tau_\mathrm{c}$ (where an asterisk again denotes the
star), which is typically $\gg\tau_\mathrm{c}$ \citep[e.g.,][]{Matsumura+10b}.
Given that $\tau_\mathrm{d}\propto1/M_\mathrm{p}$, the planets on the lower
boundary of the desert are hardly affected by this interaction; however, the
initial period distribution of the circularized orbits of large planets is
measurably modified.

\subsection{Initial Locus of Innermost Planets with Circularized Orbits}
\label{subsec:circularize}

The post-circularization locus in the ($P_\mathrm{orb}$,~$M_\mathrm{p}$) plane
of the lowest-$P_\mathrm{orb}$ planets can be obtained for both the
planet--planet scattering and Kozai migration cases from the expression for the
Roche limit (RL),
\begin{equation}
  a_\mathrm{R}=q\,(M_*/M_\mathrm{p})^{1/3}\,R_\mathrm{p}\,,\label{eq:RL}
\end{equation}
where the precise value of the coefficient $q$ depends on the planet's
structural and orbital characteristics.
Two of the most frequently cited estimates are $q=2.16$ \citep{Paczynski71} and
$q=2.7$ \citep{Guillochon+11}, but there are indications that its value in real
systems could be higher
\cite[up to $\sim3.6$--$3.8$;][]{ValsecchiRasio14,Petrovich15a}.
As it turns out, the interpretation of the sub-Jovian desert in terms of the
Roche limit also favors a comparatively large value for $q$ ($\simeq3.5$; see
Section~\ref{sec:results}).
Under the assumption that $e_0\approx1$, the circularization radius is
$\simeq2\,a_\mathrm{R}$ and thus can be written as
\begin{equation}
a_\mathrm{c,RL}\approx0.03\,\left(\frac{q}{3}\right)\left(\frac{M_*}{M_\sun}\right)^{1/3}\left(\frac{M_\mathrm{p}}{M_\mathrm{J}}\right)^{-1/3}\left(\frac{R_\mathrm{p}}{R_\mathrm{J}}\right)\,\mathrm{AU}\,.\label{eq:aRL}
\end{equation}

In the secular chaos (SC) model, the circularization radius was estimated by
\cite{WuLithwick11} as twice the pericenter distance obtained by equating the
precession rate of a planet's longitude of pericenter---due to a secular
interaction with another planet located farther out---with the orbit-averaged
precession rate due to the tidal quadrupole induced on the planet by the star.
This yields
\begin{eqnarray}
a_\mathrm{c,SC}\approx&0.03&\left(\frac{\alpha}{1/6}\right)^{-3/5}\left(\frac{M_*}{M_\sun}\right)^{2/5}\left(\frac{M_\mathrm{pert}}{M_\mathrm{J}}\right)^{-1/5}\nonumber\\&\times&\left(\frac{M_\mathrm{p}}{M_\mathrm{J}}\right)^{-1/5}\left(\frac{R_\mathrm{p}}{R_\mathrm{J}}\right)\,\mathrm{AU}\,,\label{eq:aSC}
\end{eqnarray}
where $M_\mathrm{pert}$ is the mass of the perturbing planet and $\alpha$ is the
ratio of the semimajor axes of the two planets.
For the sake of simplicity, we henceforth fix $\alpha$ at its normalization
value and treat $M_\mathrm{pert}$ as the relevant model parameter.\footnote{
An alternative possibility would have been to treat the parameter combination
$\alpha^3M_\mathrm{pert}$ as a variable.}

By combining Equations~(\ref{eq:RM}), (\ref{eq:aRL}), and~(\ref{eq:aSC}), and
using $P_\mathrm{orb}\propto{a_\mathrm{c}^{-3/2}}$, we obtain the initial (post
circularization) locus of the innermost planets for these two scenarios,
\begin{equation}
\mathrm{RL}:\left\{\begin{array}{l}P_\mathrm{orb,RL}^\mathrm{L}\propto{M_\mathrm{p}^{-1/2}}\\P_\mathrm{orb,RL}^\mathrm{S}\propto{M_\mathrm{p}^{1/4}}\end{array}\right.,\quad\mathrm{SC}:\left\{\begin{array}{l}P_\mathrm{orb,SC}^\mathrm{L}\propto{M_\mathrm{p}^{-3/5}}\\P_\mathrm{orb,SC}^\mathrm{S}\propto{M_\mathrm{p}^{3/20}}\end{array}\right.\,.
\end{equation}
It is seen that in both cases the inferred boundary has a negatively sloped
upper branch and a positively sloped lower branch.
This basic property of the observed sub-Jovian desert is a generic feature of
our proposed interpretation, arising from the different slopes in the empirical
$R_\mathrm{p}(M_\mathrm{p})$ relation for large and small planets.

\subsection{Effect of Tidally Induced Orbital Evolution}
\label{subsec:tidal}

\begin{deluxetable}{lcc}
\tablecaption{Model Parameters}
\tablehead{\colhead{Parameter}
  & \colhead{Sampled Value}       & \colhead{Distribution}}
\startdata
Number of planets
  & $5$                           &                                     \\
$P_{\mathrm{orb}}\,[\mathrm{days}]$
  & $0.5$--$50$\tablenotemark{a}  & $f(\ln{P}) \propto{P^{0.47}}$       \\
$R_\mathrm{p}\,[R_\oplus]$\; for $R_\mathrm{p}<12R_\oplus$
  & $3$--$12$\tablenotemark{a}    & $f(\ln{R})\propto{R^{-0.66}}$       \\
$M_\mathrm{p}\,[M_\oplus]$ for $R_\mathrm{p}<12R_\oplus$
  & $9$--$144$\tablenotemark{b}   & $(R_\mathrm{p}/R_\oplus)^2M_\oplus$ \\
$R_\mathrm{p}\,[M_\oplus]$\, for $R_\mathrm{p}>12R_\oplus$
  & $9$--$20$\tablenotemark{b}    & Fit to data                         \\
$M_\mathrm{p}\,[M_\oplus]$ for $R_\mathrm{p}>12R_\oplus$
  & $56$--$5620$\tablenotemark{b} & Fit to data                         \\
$t_\mathrm{arr}$ [Gyr]
  & $0.01$--$10$                  & Uniform in log time                 \\
  \enddata
\tablerefs{
  $^\mathrm{a}$~\cite{Youdin11};
  $^\mathrm{b}$~\cite{Weiss+13}.
}\label{tab:planets}
\end{deluxetable}

We calculate this effect following the procedure outlined in
\cite{MatsakosKonigl15}.
We perform Monte Carlo simulations for a population of 30,000 planetary systems,
assuming solar-type hosts ($M_*=M_\sun$, $R_*=R_\sun$) with initial rotation
periods distributed uniformly in the range $5$--$10$\,days, and drawing ages in
the range $1$--$8$\,Gyr from the empirical distribution of
\cite{WalkowiczBasri13}.
We also adopt $Q^\prime_*=10^6$ (see Section~\ref{sec:results}).
The values of $P_\mathrm{orb}$ and $R_\mathrm{p}$ are chosen from the
observationally inferred distributions presented in \cite{Youdin11}, with
$M_\mathrm{p}$ deduced from the $R_\mathrm{p}(M_\mathrm{p})$ compilation of
\cite{Weiss+13} (using their inferred power-law fit for small planets but
accounting for the observed scatter in both radius and mass for those with
$M_\mathrm{p}\gtrsim150\,M_\earth$).\footnote{
The adopted distributions of $P_{\rm orb}$ and $R_{\rm p}$ are in the form of a
single power law and thus do not capture the detailed behavior of planets in
this region of the period--mass plane.
However, since our focus is on the shape of the desert's boundary and not on
reproducing the observed density of planets, we do not consider more elaborate
distributions.}
We further assume a random distribution for the initial angle between the
stellar spin and the orbital plane, and that the distribution of planet arrival
times is uniform in $\log{t_\mathrm{arr}}$ (see Section~\ref{sec:discuss}).
These choices are summarized in Table~\ref{tab:planets}.

For each system, we integrate the evolution equations for the stellar and
orbital angular momenta, taking into account the effects of equilibrium tides
and of magnetic braking.
We assume that all orbits in a multi-planet system remain coplanar, but we
neglect planet--planet interactions.
We carry out the calculations for each of the two distributions obtained in
Section~\ref{subsec:circularize}, removing from consideration any planet drawn
with an initial semimajor axis that is less than $a_\mathrm{c,RL}$
(Equation~(\ref{eq:aRL})) or $a_\mathrm{c,SC}$ (Equation~(\ref{eq:aSC})).
Given that we do not include systems with $t_\mathrm{age}\le1$\,Gyr in our
population counts, our predicted distributions do not reveal the possible
presence of close-in giant planets that arrive at the stellar vicinity by
migration through the natal disk.
The contribution of such early-arriving planets to the observed number count is,
however, small due to their relatively rapid tidal ingestion by the star.
Nevertheless, as was noted by \cite{MatsakosKonigl15}, these HJs could have a
strong influence on the observed distribution of the angle between the stellar
spin and the orbital plane.
Based on that work, we take account of this effect by including, at time $t=0$,
a ``stranded HJ'' (SHJ) characterized by $M_\mathrm{SHJ}=0.6\,M_\mathrm{J}$,
$R_\mathrm{SHJ}=R_\mathrm{J}$, and $P_{\mathrm{SHJ}}=2\,\mathrm{days}$ in
(randomly selected) $50$\% of the modeled systems.
A giant planet of this type would undergo tidal ingestion by a G-type star on a
timescale of $\sim0.7$\,Gyr.

\section{Results}
\label{sec:results}

\begin{figure*}
  \includegraphics[width=\textwidth]{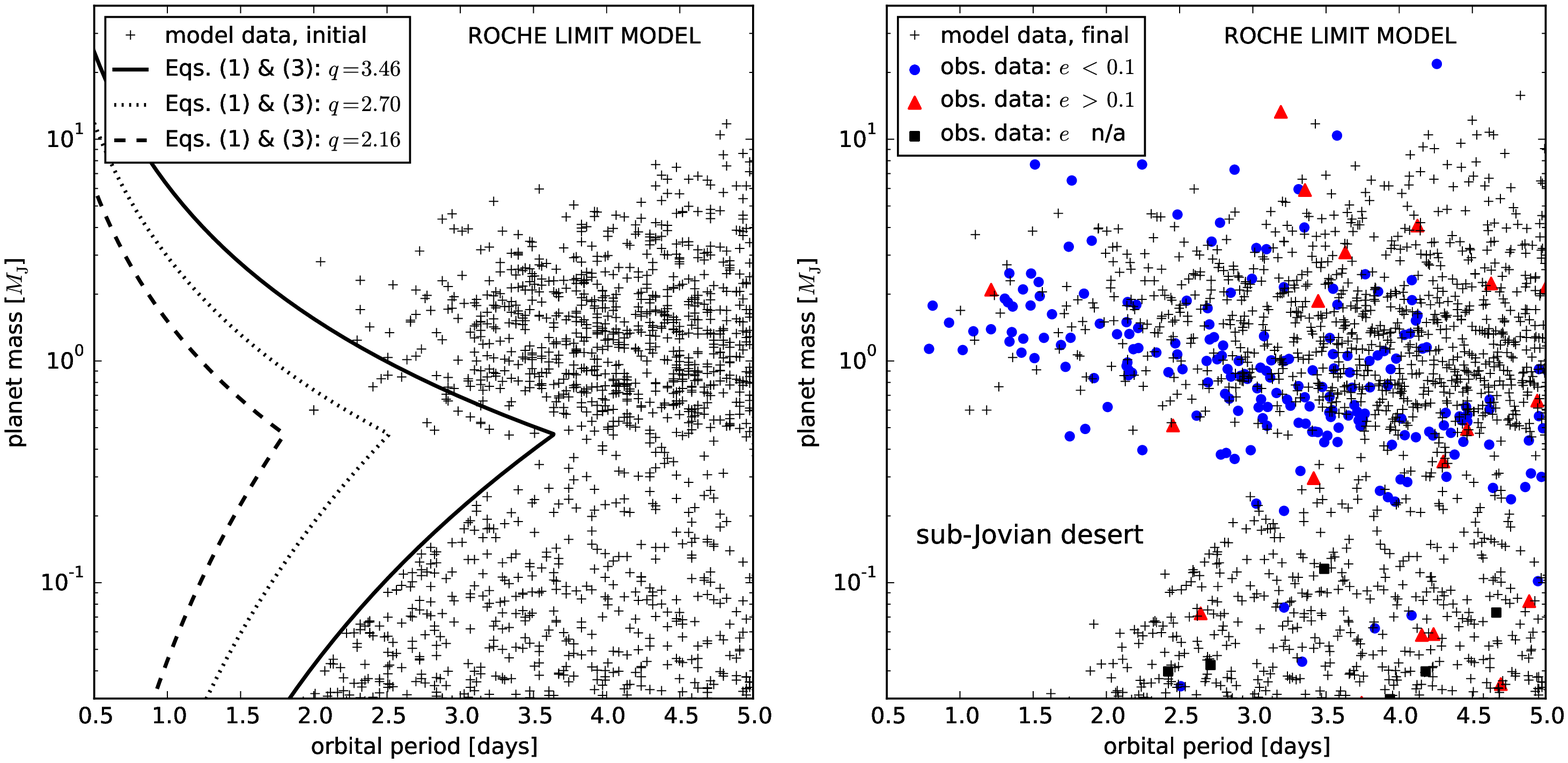}
  \includegraphics[width=\textwidth]{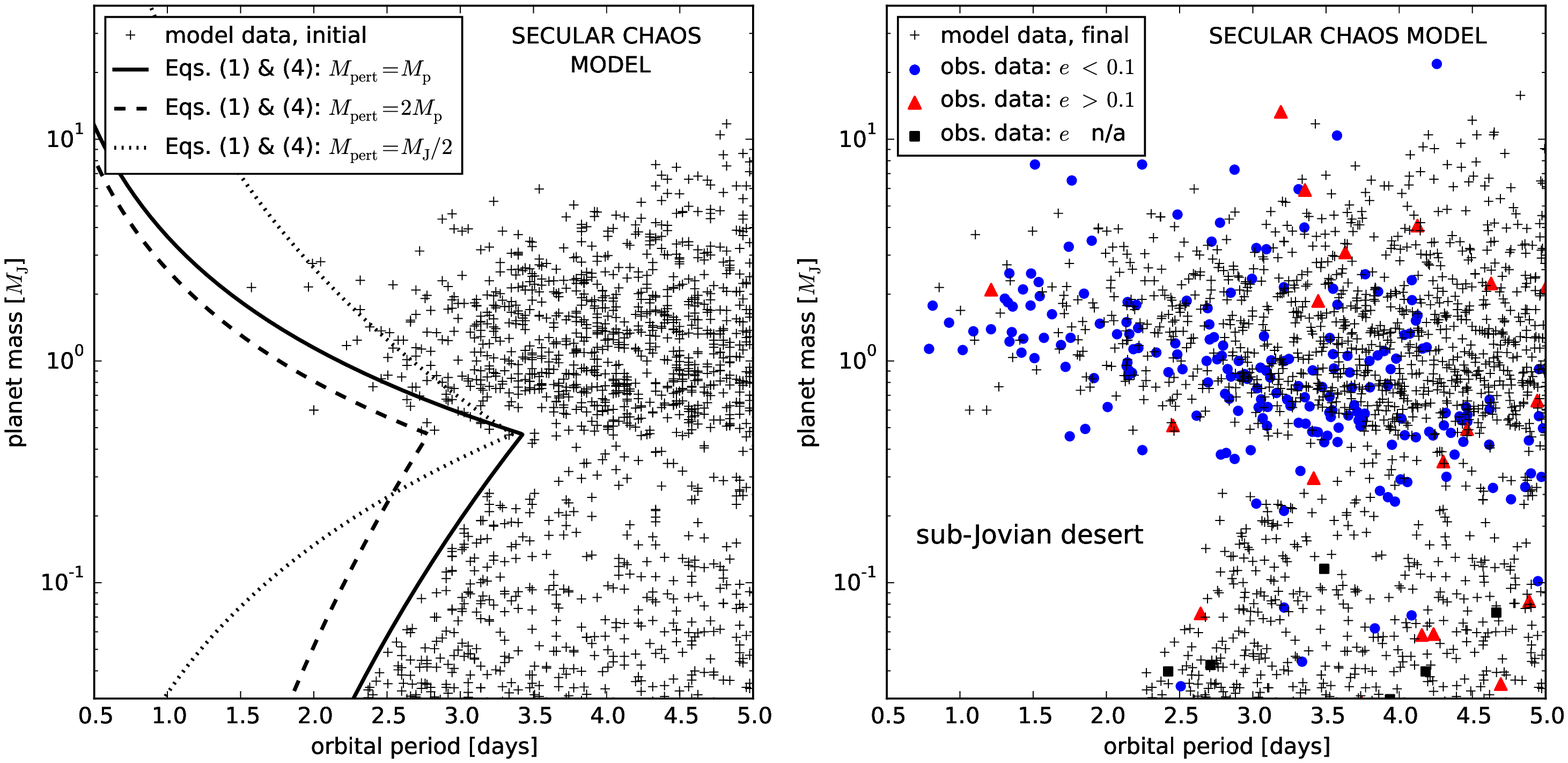}
  \caption{
    Predicted vs. observed planet distributions in the
    ($P_\mathrm{orb}$,~$\log{M_\mathrm{p}}$) plane.
    The top panels correspond to the Roche limit model for the innermost
    circularized orbits (applicable to planets arriving at the stellar vicinity
    through either scattering or Kozai migration), whereas the bottom panels
    correspond to the secular migration model.
    For each of these two cases, the left and right panels show, respectively,
    the initial distribution of planets on circular orbits and the final
    distribution obtained by calculating the effect of tidally induced orbital
    evolution.
    The initial planet distribution in the left panels was obtained by sampling
    empirical distributions of the orbital period, planetary radius, and system
    age as well as the adopted arrival-time distribution and the empirical
    $R_\mathrm{p}(M_\mathrm{p})$ relation (see Table~\ref{tab:planets}).
    Analytic predictions for the inner edge of this distribution are also shown
    in each case in the left panel for several values of the relevant model
    parameter.
    The data shown in the right panels are for all confirmed planets listed in
    \texttt{exolanets.org} \citep{Han+14} as of 2015 December 16, and are color
    coded according to their eccentricity values (when available).
    To improve the presentation, the number of displayed low-mass model planets
    was randomly reduced by $95$\% in each of the panels.\label{fig:M-P}}
\end{figure*}
Figure~\ref{fig:M-P} presents the predicted distributions in the
($P_\mathrm{orb}$,~$\log{M_\mathrm{p}}$) plane for the RL and SC models.
For the RL model we also plot the relation obtained from combining
Equations~(\ref{eq:RM}) and~(\ref{eq:aRL}) for three values of the parameter
$q$,\footnote{
The values $2.16$ and $2.70$ represent the estimates listed in
Section~\ref{subsec:circularize}, whereas $q=3.46$ corresponds to our best fit
to the desert's shape.}
whereas for the SC model we show the corresponding relation obtained from
Equations~(\ref{eq:RM}) and~(\ref{eq:aSC}) for three values of
$M_\mathrm{pert}$.\footnote{
In deriving these relations, we use the power-law scalings listed in
Equation~(\ref{eq:RM}) for small and large planets, and divide the two
populations at the value of $M_\mathrm{p}$ ($\simeq150\,M_\earth$) given in
\cite{Weiss+13}.}
It is seen that each of these scenarios can reproduce the observed shape of the
sub-Jovian desert's boundary quite well.
The initial circularization locus accounts for the basic ``bird's beak''
configuration of the boundary, but only the shape of its low-mass segment (along
which orbital evolution effects are negligible) and the location of the vertex
(where the upper and lower segments intersect) are preserved in the final
distribution.
The mass coordinate of the vertex is fixed by the transition point
($M_\mathrm{p}\approx150\,M_\earth$) of the empirical mass--radius relation and
is independent of the details of the underlying planet migration model; the fact
that it matches the observations so accurately is a strong indication that this
aspect of the small/large planets dichotomy plays a key role in determining the
desert's shape.
On the other hand, the plots in the left panels of Figure~\ref{fig:M-P}
demonstrate that the period coordinate of the vertex is sensitive to the value
of the relevant model parameter.
In each case, however, the inferred best-fit value has a plausible magnitude.
In the RL picture, the comparatively large indicated value of $q$ ($\simeq3.5$)
is consistent with other recent determinations (e.g.,
\citealt{ValsecchiRasio14}, who derived the initial locations of HJs detected
inside the Roche limit; and \citealt{Petrovich15a}, who investigated the
formation of HJs by Kozai migration).
In the case of secular chaos, the inference that
$M_\mathrm{pert}\approx{M_\mathrm{p}}$ is also eminently reasonable.
For Jupiter-mass planets, the excitation of secular interactions would be much
weaker if the companion planet's mass were measurably less than that of the
innermost planet \citep[e.g.,][]{WuLithwick11}, whereas for SE planets,
comparable-mass objects are the most frequent companions and are thus the most
likely to act as perturbers.
Note that the value of the model parameter ($q$ or $M_\mathrm{pert}$) also
affects the predicted shape of the lower branch of the desert's boundary, which
further constrains the choice of the best match.

The shape of the upper boundary is largely independent of the details of the
initial distribution of high-mass planets, and is determined mainly by the
ensuing orbital evolution.
It is well reproduced---in the context of the equilibrium tidal interaction
model that we employ---with a stellar tidal quality factor $Q^\prime_*\sim10^6$,
which is consistent with previous inferences from modeling the $P_\mathrm{orb}$
distribution of HJs \citep[e.g.,][]{TeitlerKoenigl14,EssickWeinberg16}.
It is noteworthy that the two alternative models originally proposed to explain
the spatial distribution of HJs---circularization of highly eccentric orbits and
tidal exchange of angular momentum with the star---are both found to be relevant
in the context of the current interpretation when the detailed distribution in
the ($P_\mathrm{orb}$,~$M_\mathrm{p}$) plane is taken into account.

\section{Discussion}
\label{sec:discuss}

Our proposed interpretation of the sub-Jovian desert's shape---that it is a
natural consequence of the orbital circularization of planets that arrive at the
stellar vicinity on high-eccentricity orbits and of their subsequent tidal
angular-momentum exchange with the star---is consistent with the finding that
the eccentricity distribution of giant planets broadens with increasing period
\citep[e.g.,][]{WinnFabrycky15} and with the evidence for ongoing orbital
evolution of the innermost HJs \citep[e.g.,][]{ValsecchiRasio14,
TeitlerKoenigl14,EssickWeinberg16}.
The latter result, in turn, supports the view that the observed HJs are mostly
late-arriving planets rather than the product of migration in the protoplanetary
disk.
The high-eccentricity migration scenario, in which the ingoing planet is placed
on a high-$e_0$ orbit through gravitational interaction with one or more massive
bodies (other planets or a binary star), is consistent with the indicated high
occurrence rate of planetary and stellar companions in systems that harbor HJs
\citep[e.g.][]{Knutson+14,Ngo+15}.
We considered three possible pathways to such an outcome---planet-planet
scattering, Kozai migration, and secular drift---and inferred that in principle
they could all play a role.
However, a more detailed examination is required to determine the actual
contribution of each of these processes.
One pertinent question is whether the influence of the process extends to late
times; for example, secular chaos is inherently a long-lasting interaction, for
which the expected distribution of arrival times is approximately uniform in
$\log{t_\mathrm{arr}}$ (Y.~Lithwick, personal communication), whereas scattering
likely only plays a role at early times
\cite[$\lesssim10^8$\,yr; e.g.,][]{Chatterjee+08,JuricTremaine08}.
Another relevant question concerns the rate of eccentricity growth in the
high-$e$ migration model; for example, \cite{WuLithwick11} argued that the
observed $\sim3$-day pileup of HJs is more likely to have been caused by a slow
process such as Kozai or secular migration than by sudden scattering events
\cite[see also][]{Nagasawa+08}.
Further studies are needed to fully address this issue.

One attractive feature of the proposed interpretation is that it can
simultaneously reproduce both the upper and the lower segments of the desert's
boundary.
An alternative possibility for the low-mass branch is that it arises from the
in-situ formation of close-in SEs \citep{Mazeh+16}, which has been recently
explored in the literature \citep[e.g.,][]{LeeChiang16}.
However, arguments have also been given in favor of formation at larger
distances (up to a few AU) and subsequent inward migration
\citep[e.g.,][]{Schlichting14,InamdarSchlichting15}.
Our model for the sub-Jovian desert is consistent with the latter picture.

The scenario considered in this work, which involves late-arriving planets,
complements the SHJs model presented in \cite{MatsakosKonigl15}, which is
concerned with early-arriving HJs.
That model explains the good alignment exhibited by a significant fraction of
HJs around cool (G type) stars---as well as the good alignment inferred for more
distant planets around such stars---vs. the broad range of obliquities exhibited
by HJs around hot (F type) stars.
As we noted in Section~\ref{sec:model}, the earlier population of HJs would not
affect the observed properties of the sub-Jovian desert because of the
relatively short SHJ ingestion time.
However, consistency with the SHJs scenario requires the orbital plane of any
late-arriving planet to roughly coincide with that of the natal disk.
This additional constraint on the high-eccentricity migration model need not,
however, be too difficult to fulfill.
For example, \cite{Matsumura+10a} found that only $\sim15\%$ of planets in a
3-planet system that emerges out of a gas disk have orbital inclinations
$>10^\circ$, and \cite{LithwickWu14} discovered that, even if larger initial
inclinations are allowed, $60\%$ of the HJs formed through secular migration
involving a 3-planet system have projected obliquities $<10^\circ$.
Furthermore, \cite{Petrovich15b} demonstrated the feasibility of producing HJs
through coplanar high-$e$ migration (a secular process involving 2 planets).

\acknowledgements
We are grateful to Tsevi Mazeh for alerting us to the ``bird's beak'' shape of
the sub-Jovian desert's boundary.
We thank him as well as Dan Fabrycky, Yoram Lithwick, Leslie Rogers, and the
referee for helpful input.
This work was supported in part by NASA ATP grant NNX13AH56G.


\begin{thebibliography}

\bibitem[{{Batygin} {et~al.}(2016){Batygin}, {Bodenheimer}, \&
  {Laughlin}}]{Batygin+16}
{Batygin}, K., {Bodenheimer}, P.~H., \& {Laughlin}, G.~P. 2016, ApJ, submitted,
  arXiv:1511.09157

\bibitem[{{Beaug{\'e}} \& {Nesvorn{\'y}}(2013)}]{BeaugeNesvorny13}
{Beaug{\'e}}, C., \& {Nesvorn{\'y}}, D. 2013, \apj, 763, 12

\bibitem[{{Ben{\'{\i}}tez-Llambay} {et~al.}(2011){Ben{\'{\i}}tez-Llambay},
  {Masset}, \& {Beaug{\'e}}}]{Benitez-Llambay+11}
{Ben{\'{\i}}tez-Llambay}, P., {Masset}, F., \& {Beaug{\'e}}, C. 2011, \aap,
  528, A2

\bibitem[{{Boley} {et~al.}(2016){Boley}, {Granados Contreras}, \&
  {Gladman}}]{Boley+16}
{Boley}, A.~C., {Granados Contreras}, A.~P., \& {Gladman}, B. 2016, \apjl, 817,
  L17

\bibitem[{{Chatterjee} {et~al.}(2008){Chatterjee}, {Ford}, {Matsumura}, \&
  {Rasio}}]{Chatterjee+08}
{Chatterjee}, S., {Ford}, E.~B., {Matsumura}, S., \& {Rasio}, F.~A. 2008, \apj,
  686, 580

\bibitem[{{Col{\'o}n} {et~al.}(2015){Col{\'o}n}, {Morehead}, \&
  {Ford}}]{Colon+15}
{Col{\'o}n}, K.~D., {Morehead}, R.~C., \& {Ford}, E.~B. 2015, \mnras, 452, 3001

\bibitem[{{Cumming} {et~al.}(2008){Cumming}, {Butler}, {Marcy}, {Vogt},
  {Wright}, \& {Fischer}}]{Cumming+08}
{Cumming}, A., {Butler}, R.~P., {Marcy}, G.~W., {et~al.} 2008, \pasp, 120, 531

\bibitem[{{Essick} \& {Weinberg}(2016)}]{EssickWeinberg16}
{Essick}, R., \& {Weinberg}, N.~N. 2016, \apj, 816, 18

\bibitem[{{Fabrycky} \& {Tremaine}(2007)}]{FabryckyTremaine07}
{Fabrycky}, D., \& {Tremaine}, S. 2007, \apj, 669, 1298

\bibitem[{{Ford} \& {Rasio}(2006)}]{FordRasio06}
{Ford}, E.~B., \& {Rasio}, F.~A. 2006, \apjl, 638, L45

\bibitem[{{Fressin} {et~al.}(2013){Fressin}, {Torres}, {Charbonneau}, {Bryson},
  {Christiansen}, {Dressing}, {Jenkins}, {Walkowicz}, \&
  {Batalha}}]{Fressin+13}
{Fressin}, F., {Torres}, G., {Charbonneau}, D., {et~al.} 2013, \apj, 766, 81

\bibitem[{{Gaudi} {et~al.}(2005){Gaudi}, {Seager}, \&
  {Mallen-Ornelas}}]{Gaudi+05}
{Gaudi}, B.~S., {Seager}, S., \& {Mallen-Ornelas}, G. 2005, \apj, 623, 472

\bibitem[{{Goldreich} \& {Soter}(1966)}]{GoldreichSoter66}
{Goldreich}, P., \& {Soter}, S. 1966, Icarus, 5, 375

\bibitem[{{Guillochon} {et~al.}(2011){Guillochon}, {Ramirez-Ruiz}, \&
  {Lin}}]{Guillochon+11}
{Guillochon}, J., {Ramirez-Ruiz}, E., \& {Lin}, D. 2011, \apj, 732, 74

\bibitem[{{Han} {et~al.}(2014){Han}, {Wang}, {Wright}, {Feng}, {Zhao},
  {Fakhouri}, {Brown}, \& {Hancock}}]{Han+14}
{Han}, E., {Wang}, S.~X., {Wright}, J.~T., {et~al.} 2014, \pasp, 126, 827

\bibitem[{{Howard} {et~al.}(2012){Howard}, {Marcy}, {Bryson}, {Jenkins},
  {Rowe}, {Batalha}, {Borucki}, {Koch}, {Dunham}, {Gautier}, {Van Cleve},
  {Cochran}, {Latham}, {Lissauer}, {Torres}, {Brown}, {Gilliland}, {Buchhave},
  {Caldwell}, {Christensen-Dalsgaard}, {Ciardi}, {Fressin}, {Haas}, {Howell},
  {Kjeldsen}, {Seager}, {Rogers}, {Sasselov}, {Steffen}, {Basri},
  {Charbonneau}, {Christiansen}, {Clarke}, {Dupree}, {Fabrycky}, {Fischer},
  {Ford}, {Fortney}, {Tarter}, {Girouard}, {Holman}, {Johnson}, {Klaus},
  {Machalek}, {Moorhead}, {Morehead}, {Ragozzine}, {Tenenbaum}, {Twicken},
  {Quinn}, {Isaacson}, {Shporer}, {Lucas}, {Walkowicz}, {Welsh}, {Boss},
  {Devore}, {Gould}, {Smith}, {Morris}, {Prsa}, {Morton}, {Still}, {Thompson},
  {Mullally}, {Endl}, \& {MacQueen}}]{Howard+12}
{Howard}, A.~W., {Marcy}, G.~W., {Bryson}, S.~T., {et~al.} 2012, \apjs, 201, 15

\bibitem[{{Inamdar} \& {Schlichting}(2015)}]{InamdarSchlichting15}
{Inamdar}, N.~K., \& {Schlichting}, H.~E. 2015, \mnras, 448, 1751

\bibitem[{{Jackson} {et~al.}(2009){Jackson}, {Barnes}, \&
  {Greenberg}}]{Jackson+09}
{Jackson}, B., {Barnes}, R., \& {Greenberg}, R. 2009, \apj, 698, 1357

\bibitem[{{Juri{\'c}} \& {Tremaine}(2008)}]{JuricTremaine08}
{Juri{\'c}}, M., \& {Tremaine}, S. 2008, \apj, 686, 603

\bibitem[{{Knutson} {et~al.}(2014){Knutson}, {Fulton}, {Montet}, {Kao}, {Ngo},
  {Howard}, {Crepp}, {Hinkley}, {Bakos}, {Batygin}, {Johnson}, {Morton}, \&
  {Muirhead}}]{Knutson+14}
{Knutson}, H.~A., {Fulton}, B.~J., {Montet}, B.~T., {et~al.} 2014, \apj, 785,
  126

\bibitem[{{Kurokawa} \& {Nakamoto}(2014)}]{KurokawaNakamoto14}
{Kurokawa}, H., \& {Nakamoto}, T. 2014, \apj, 783, 54

\bibitem[{{Lee} \& {Chiang}(2016)}]{LeeChiang16}
{Lee}, E.~J., \& {Chiang}, E. 2016, \apj, 817, 90

\bibitem[{{Lithwick} \& {Wu}(2014)}]{LithwickWu14}
{Lithwick}, Y., \& {Wu}, Y. 2014, Proceedings of the National Academy of
  Science, 111, 12610

\bibitem[{{Matsakos} \& {K{\"o}nigl}(2015)}]{MatsakosKonigl15}
{Matsakos}, T., \& {K{\"o}nigl}, A. 2015, \apjl, 809, L20

\bibitem[{{Matsumura} {et~al.}(2010{\natexlab{a}}){Matsumura}, {Peale}, \&
  {Rasio}}]{Matsumura+10b}
{Matsumura}, S., {Peale}, S.~J., \& {Rasio}, F.~A. 2010{\natexlab{a}}, \apj,
  725, 1995

\bibitem[{{Matsumura} {et~al.}(2010{\natexlab{b}}){Matsumura}, {Thommes},
  {Chatterjee}, \& {Rasio}}]{Matsumura+10a}
{Matsumura}, S., {Thommes}, E.~W., {Chatterjee}, S., \& {Rasio}, F.~A.
  2010{\natexlab{b}}, \apj, 714, 194

\bibitem[{{Mazeh} {et~al.}(2016){Mazeh}, {Holczer}, \& {Faigler}}]{Mazeh+16}
{Mazeh}, T., {Holczer}, T., \& {Faigler}, S. 2016, A\&A, submitted

\bibitem[{{Nagasawa} {et~al.}(2008){Nagasawa}, {Ida}, \&
  {Bessho}}]{Nagasawa+08}
{Nagasawa}, M., {Ida}, S., \& {Bessho}, T. 2008, \apj, 678, 498

\bibitem[{{Ngo} {et~al.}(2015){Ngo}, {Knutson}, {Hinkley}, {Crepp}, {Bechter},
  {Batygin}, {Howard}, {Johnson}, {Morton}, \& {Muirhead}}]{Ngo+15}
{Ngo}, H., {Knutson}, H.~A., {Hinkley}, S., {et~al.} 2015, \apj, 800, 138

\bibitem[{{Paczy{\'n}ski}(1971)}]{Paczynski71}
{Paczy{\'n}ski}, B. 1971, \araa, 9, 183

\bibitem[{{Petrovich}(2015{\natexlab{a}})}]{Petrovich15b}
{Petrovich}, C. 2015{\natexlab{a}}, \apj, 805, 75

\bibitem[{{Petrovich}(2015{\natexlab{b}})}]{Petrovich15a}
---. 2015{\natexlab{b}}, \apj, 799, 27

\bibitem[{{Rogers}(2015)}]{Rogers15}
{Rogers}, L.~A. 2015, \apj, 801, 41

\bibitem[{{Schlichting}(2014)}]{Schlichting14}
{Schlichting}, H.~E. 2014, \apjl, 795, L15

\bibitem[{{Szab{\'o}} \& {Kiss}(2011)}]{SzaboKiss11}
{Szab{\'o}}, G.~M., \& {Kiss}, L.~L. 2011, \apjl, 727, L44

\bibitem[{{Teitler} \& {K{\"o}nigl}(2014)}]{TeitlerKoenigl14}
{Teitler}, S., \& {K{\"o}nigl}, A. 2014, \apj, 786, 139

\bibitem[{{Valsecchi} {et~al.}(2015){Valsecchi}, {Rappaport}, {Rasio},
  {Marchant}, \& {Rogers}}]{Valsecchi+15}
{Valsecchi}, F., {Rappaport}, S., {Rasio}, F.~A., {Marchant}, P., \& {Rogers},
  L.~A. 2015, \apj, 813, 101

\bibitem[{{Valsecchi} \& {Rasio}(2014)}]{ValsecchiRasio14}
{Valsecchi}, F., \& {Rasio}, F.~A. 2014, \apjl, 787, L9

\bibitem[{{Valsecchi} {et~al.}(2014){Valsecchi}, {Rasio}, \&
  {Steffen}}]{Valsecchi+14}
{Valsecchi}, F., {Rasio}, F.~A., \& {Steffen}, J.~H. 2014, \apjl, 793, L3

\bibitem[{{Walkowicz} \& {Basri}(2013)}]{WalkowiczBasri13}
{Walkowicz}, L.~M., \& {Basri}, G.~S. 2013, \mnras, 436, 1883

\bibitem[{{Weiss} {et~al.}(2013){Weiss}, {Marcy}, {Rowe}, {Howard}, {Isaacson},
  {Fortney}, {Miller}, {Demory}, {Fischer}, {Adams}, {Dupree}, {Howell},
  {Kolbl}, {Johnson}, {Horch}, {Everett}, {Fabrycky}, \& {Seager}}]{Weiss+13}
{Weiss}, L.~M., {Marcy}, G.~W., {Rowe}, J.~F., {et~al.} 2013, \apj, 768, 14

\bibitem[{{Winn} \& {Fabrycky}(2015)}]{WinnFabrycky15}
{Winn}, J.~N., \& {Fabrycky}, D.~C. 2015, \araa, 53, 409

\bibitem[{{Wu} \& {Lithwick}(2011)}]{WuLithwick11}
{Wu}, Y., \& {Lithwick}, Y. 2011, \apj, 735, 109

\bibitem[{{Wu} \& {Murray}(2003)}]{WuMurray03}
{Wu}, Y., \& {Murray}, N. 2003, \apj, 589, 605

\bibitem[{{Youdin}(2011)}]{Youdin11}
{Youdin}, A.~N. 2011, \apj, 742, 38

\bibitem[{{Zucker} \& {Mazeh}(2002)}]{ZuckerMazeh02}
{Zucker}, S., \& {Mazeh}, T. 2002, \apjl, 568, L113

\end{thebibliography}
\end{document}